\documentclass[%
prl, nofootinbib, reprint,
 amsmath,amssymb,
 aps,
]{revtex4-1}

\usepackage{cancel}
\usepackage{graphicx}
\usepackage{dcolumn}
\usepackage{bm}
\usepackage{soul}
\usepackage[normalem]{ulem}

\usepackage{amssymb}
\usepackage{amsmath}
\usepackage{hyperref}
\usepackage{color}
\usepackage{rotate}
\usepackage{fancyhdr}

\def\be{\begin{equation}}
\def\ee{\end{equation}}
\def\bea{\begin{eqnarray}}
\def\eea{\end{eqnarray}}
\def\nn{\nonumber}

\def\hksqrt{\mathpalette\DHLhksqrt}
\def\DHLhksqrt#1#2{\setbox0=\hbox{$#1\sqrt{#2\,}$}\dimen0=\ht0
\advance\dimen0-0.2\ht0
\setbox2=\hbox{\vrule height\ht0 depth -\dimen0}%
{\box0\lower0.4pt\box2}}

\definecolor{darkred}{RGB}{175,0,0}

\begin{document}

\preprint{APS/123-QED}

\title{Detecting the integrated Sachs-Wolfe effect with high-redshift 21-cm surveys}

\author{Alvise Raccanelli}
\author{Ely Kovetz}
\author{Liang Dai}
\author{Marc Kamionkowski}
\affiliation{%
 Department of Physics \& Astronomy, Johns Hopkins University, 3400 N. Charles St., Baltimore, MD 21218, USA \\
}%

\date{\today}

\begin{abstract}
We investigate the possibility to detect the integrated Sachs-Wolfe (ISW) effect by cross-correlating 21-cm surveys at high redshifts with galaxies, in a way similar to the usual CMB-galaxy cross-correlation. The high-redshift 21-cm signal is dominated by CMB photons that travel freely without interacting with the intervening matter, and hence its late-time ISW signature should correlate extremely well with that of the CMB at its peak frequencies. Using the 21-cm temperature brightness instead of the CMB would thus be a further check of the detection of the ISW effect, measured by different instruments at different frequencies and suffering from different systematics. We also study the ISW effect on the photons that are scattered by HI clouds. 
We show that a detection of the unscattered photons is achievable with planned radio arrays, while one using scattered photons will require advanced radio interferometers, either an extended version of the planned Square Kilometre Array or futuristic experiments such as a lunar radio array.
\end{abstract}

\maketitle


The integrated Sachs-Wolfe (ISW) effect~\cite{sachs67, crittenden96, nishizawa14} is a gravitational redshift due to the evolution of the gravitational potential as photons pass through matter
under- and over-densities in their path from the last scattering surface to us. In an Einstein-de Sitter universe, the blueshift of a
photon falling into a well is cancelled by the redshift as it climbs out, giving a zero effect. In the presence of a time variation
of the local gravitational potential, due to a dark energy component or a deviation from Einstein's theory of gravity, however,
potential wells are modified and photons will experience a blue- or red-shift, leading to a net change in photon temperature
which accumulates along the photon path. This effect translates into temperature anisotropies proportional to the variation of the
gravitational potentials.
The ISW effect has been detected~\cite{nolta04, pietrobon06, ho08, giannantonio08a, raccanelli08, giannantonio12, planckisw, planckisw2015} through cross-correlation of CMB maps at ${\rm GHz}$-frequencies with galaxy surveys and used to constrain cosmological parameters~\cite{giannantonio08dgp, massardi10, bertacca11, Raccanelli:2014ISW}.

\begin{figure}[b!]
\includegraphics[width=0.99\columnwidth]{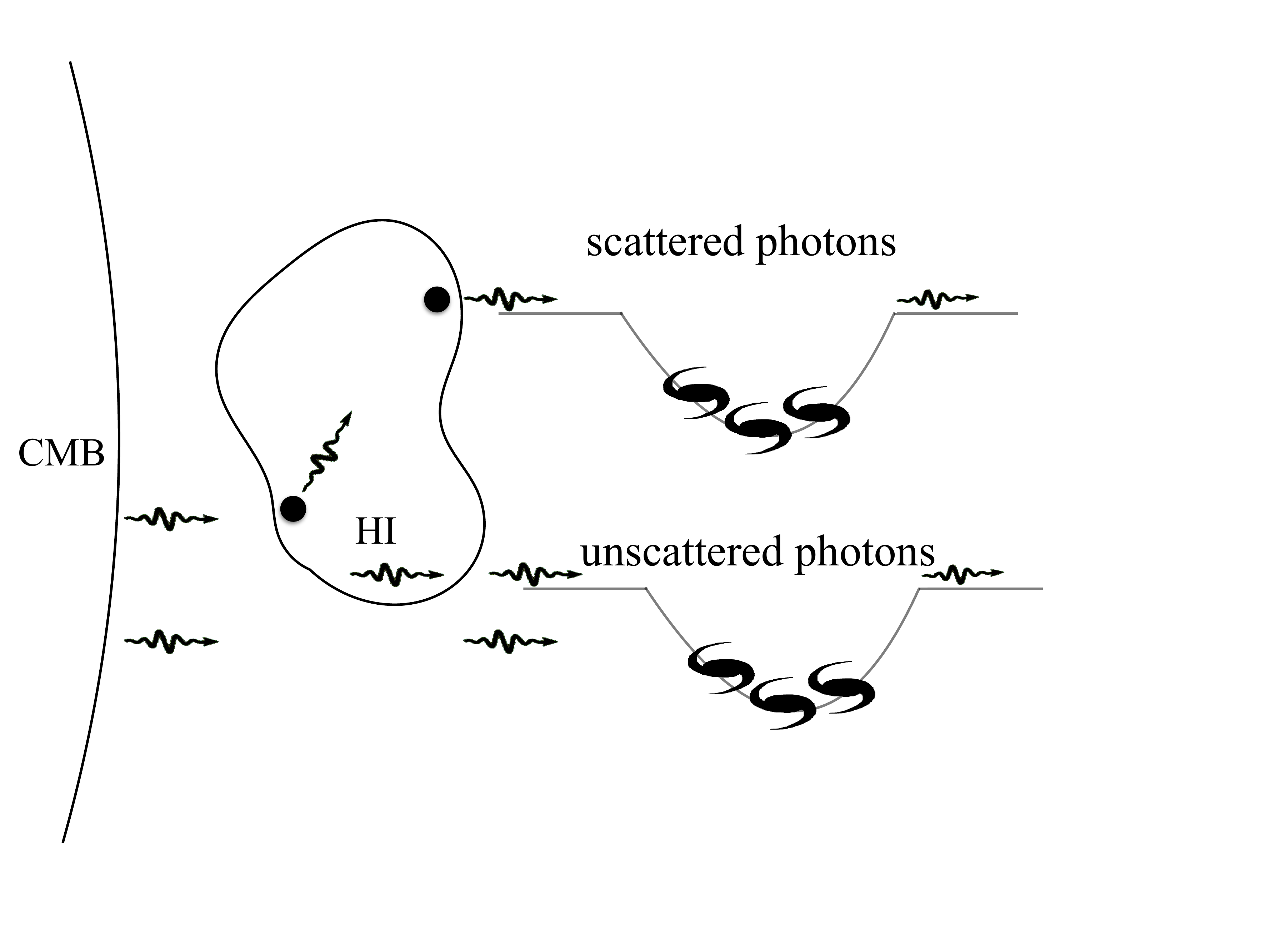}
\caption{
{\it Illustration:} The radiative transfer of CMB photons through neutral hydrogen gas clouds induces fluctuations at 21-cm frequencies (due to absorption or emission, depending on the relative temperatures of the inter-galactic medium and the CMB) which are the target of many ongoing and planned experiments. The majority of the signal, however, is comprised of unscattered CMB photons at the Rayleigh-Jeans tail of its blackbody spectrum. All these photons later undergo line-of-sight blue- or red- shifting as they travel through evolving gravitational potential wells (e.g.\ during $\Lambda$-domination). This integrated Sachs-Wolfe contribution to the fluctuations at 21-cm frequencies can then be efficiently extracted by cross-correlating the maps with low-redshift galaxy surveys.}
\label{fig:illustration}
\end{figure}

In a similar way, the ``21-cm background" at high-redshift, which is sourced by CMB photons in the Rayleigh-Jeans tail of the blackbody spectrum that travel through neutral hydrogen clouds, will experience an ISW effect from the evolution of gravitational potential wells (see Fig.~\ref{fig:illustration}).
Stripped of its unscattered CMB contribution, the 21-cm signal represents a unique source of cosmological information from the epoch of reionization (EOR) and the dark ages~\cite{Furlanetto, Pritchard:2011}; measuring at different frequencies, the large redshift volume of these two epochs can be used to probe the spatial distribution of neutral hydrogen~\cite{Zaldarriaga, Loeb, Lewis}.
The 3D power spectrum and imaging of the 21-cm brightness temperature fluctuations can also be used for various other cosmological measurements (see e.g.~\cite{kovetz12, mao13}). Given that mapping the 21-cm signal is the target of several ongoing and near-future ground-based experiments such as the SKA~\cite{SKA}, as well as more futuristic experiments on the far side of the Moon~\cite{lunar, lunar2} or space-based radio arrays like RadioAstron~\cite{radioastron}, DARE~\cite{dare}, DALI~\cite{dali}, OLFAR~\cite{olfar}, DARIS~\cite{daris}, SURO~\cite{suro}, PARIS~\cite{paris}, that promise to reach very high accuracy, it is worthwhile to investigate how to maximize the scientific output of such measurements. In this {\it Letter} we briefly present the formalism for studying the ISW effect on the 21-cm signal and the experimental challenges for its detection, using the cross-correlation of low-redshift tracers of matter (galaxies) with maps of brightness temperature fluctuations at 21-cm frequencies corresponding to high redshifts (see illustration in Fig.~\ref{fig:illustration}).

The dominant signal in these maps is of unscattered CMB photons, and therefore its late-time ISW signature will be very highly correlated with the signature at the peak CMB frequencies mentioned above. A complementing measurement at 21-cm frequencies would be promising as it represents an independent detection of the ISW effect, measured with different instruments and contaminated by different foregrounds. As the 21-cm background is set to be observed across a vast redshift range by upcoming experiments, we show that there should be ample signal-to-noise for this detection.
There is, however, another feature that could be sought after, which is the ISW effect on those CMB photons that \textit{do} interact with the neutral hydrogen clouds at high redshifts. Assuming the CMB fluctuations are efficiently subtracted from the 21-cm maps, we demonstrate that this signal can potentially be detected in the data as well, although most-likely requiring futuristic experiments.


We parameterize potential perturbations in the conformal Newtonian (cN) gauge
\bea
ds^2 & = & a^2(\eta) \left[ - (1+2\Psi) d\eta^2 + \left( 1 - 2\Phi \right) \delta_{ij} dx^i dx^j \right].
\eea
The (fractional) shift in the observed photon frequency is given by
\bea
\label{eq:eq-2}
\frac{\delta\nu}{\nu} = \left( \Psi_s - \Psi_o \right) + \int^s_o\,d\chi \left( \dot{\Psi} + \dot{\Phi} \right).
\eea 
The right hand side is frequency-independent. The first term, known as the Sachs-Wolfe term, is the potential difference between the source location ${}_s$ and the observer's location ${}_o$. The second term is an ISW line-of-sight integral from the observer to the source, with $\chi$ being the comoving radial distance. In this work we focus on the second term. There is also a second order effect, called Rees-Sciama~\cite{rees68}, due to the evolution of clustering while the photon is traversing the potential well, that happens even in the absence of dark energy, but this is typically orders of magnitude smaller than the ISW effect, so we will neglect it.

The observed 21-cm signature at frequency $\nu$ will correspond to frequency $\nu-\delta \nu$ without the ISW effect. By conservation of photon number, the observed occupation number (per polarization state) $f(\nu)$ would be
\bea
f(\nu) = f_0(\nu - \delta \nu) = f_0(\nu) \left[ 1 - \frac{\partial \ln f}{\partial\ln \nu} \frac{\delta\nu}{\nu} \right],
\eea
where $f_0(\nu)$ is the occupation number without ISW. The fractional change is therefore given by
\bea
\frac{\delta f}{f} = - \frac{\partial \ln f}{\partial\ln \nu} \frac{\delta\nu}{\nu}.
\eea
At a given observed frequency $\nu$, the specific intensity $I(\nu)$ (for both polarization states) is related to the occupation number by $I(\nu) d\nu = 2 f(\nu) \nu^2 d\nu$. We therefore find a fractional change (note that $\nu$ is fixed)
\bea
\frac{\delta I}{I} = \frac{\delta f}{f} = - \frac{\partial \ln f}{\partial\ln \nu} \frac{\delta\nu}{\nu}.
\eea

The 21-cm signature is sourced by photons from the CMB background that traverse neutral hydrogen regions before reaching the observer. If interaction with neutral hydrogen clouds is neglected, the occupation number is given by a blackbody spectrum, but in the Rayleigh-Jeans regime
\bea
f_R = \frac{T_{R0}}{\nu},
\eea
characterized by a thermodynamic temperature $T_{R0}=2.725~$K (at present redshift). It then follows that $-\partial\ln f/\partial\ln \nu=1$, and therefore
\bea
\frac{\delta I}{I} = \frac{\delta\nu}{\nu} = \left[ \int^s_o\,d\chi \left( \dot{\Psi} + \dot{\Phi} \right) \right].
\eea
Since we are in the Rayleigh-Jeans limit, the specific intensity is proportional to the temperature, and we reproduce the ISW contribution to CMB temperature anisotropies
\bea
\frac{\delta T_{R}}{T_{R0}} = \frac{\delta I}{I} = \left[ \int^s_o\,d\chi \left( \dot{\Psi} + \dot{\Phi} \right) \right].
\eea

If the interaction between the CMB and neutral hydrogen is accounted for, the occupation number $f(\nu)$ will be distorted
\bea
f(\nu) = f_{R}(\nu) + \frac{T_{\rm 21}}{\nu} = f_R \left[ 1 + \frac{T_{\rm 21}}{f_R\, \nu} \right],
\eea
where the mean brightness temperature $T_{21}$ is given by
\begin{equation}
T_{21}(\nu) \approx \frac{T_S(z)-T_R(z)}{1+z}\tau_{21}(z) \,
\end{equation}
corresponding to the redshift $z=z(\nu)$ of CMB-atom interaction. Here $\tau_{\rm 21}$ is the optical depth for the hyperfine transition~\cite{pillepich}
\begin{equation}
\tau_{21} \approx 0.025 \frac{T_{R0}}{T_s} \frac{\Omega_b \, h}{0.035} \left( \frac{\Omega_m}{0.27} \right)^{-1/2} \left(\frac{1+z}{51} \right)^{1/2} \, .
\end{equation}
Thus, the intensity fluctuation is given by
\bea
\frac{\delta I}{I} & = & \left[ 1 - \frac{\partial}{\partial\ln \nu} \left( \frac{T_{\rm 21}}{f_R\, \nu} \right) \right] \left[ \int^s_o\,d\chi \left( \dot{\Psi} + \dot{\Phi} \right) \right]  \nn\\
& = & \left[ 1 - \frac{T_{\rm 21}}{ T_{R0}} \frac{\partial\ln T_{\rm 21}}{\partial\ln \nu} \right] \left[ \int^s_o\,d\chi \left( \dot{\Psi} + \dot{\Phi} \right) \right] \, .
\label{eq:main}
\eea
Eq.~(\ref{eq:main}) is the main equation of this paper, and includes the two signatures we are targeting. The second term in the parentheses is suppressed by $T_{\rm 21}/T_{R0} \sim 10^{-2} - 10^{-1}$, and is therefore generically much smaller than unity. As a first approximation, we can choose to neglect it and consider first the detection of the ISW effect on the unscattered CMB photons, represented by the first term. Then, assuming the unscattered CMB fluctuations can be removed using external CMB datasets, we can estimate if the ISW effect on the scattered photons can be detected as well. For this purpose, the ideal redshift would be $z\lesssim30$, where the derivative $\partial \ln T_{\rm 21}/\partial \ln \nu$ can turn out to be large (see e.g.\ Figs. 4,\! 7 of~\cite{Pritchard:2011}). This redshift corresponds to the very end of the dark ages and might be within reach of the SKA. The exact value of the second-term coefficient is highly model-dependent, though (mainly dictated by the X-ray and Lyman-$\alpha$ emissivity), and we choose a fiducial (optimistic) value of 0.3 for the purpose of examining its detectability.
Since photons from $z\simeq1100$ pass through an epoch with some residual radiation domination, they undergo an additional early ISW redshift not experienced by those from $z\simeq30$.  There will therefore perhaps be subtle differences between the temperature fluctuation in the scattered and unscattered photons.  We leave a more detailed discussion of this effect to future work.

%

The ISW effect can contribute significantly to the temperature fluctuations on large angular scales, but it enhances only
the low multipoles, and it is generally smaller than the temperature power spectrum, both at the CMB peak frequencies and at those corresponding to high redshift 21-cm. Typically the ISW signal is weak and cross-correlation with tracers of the density field is necessary to bring it to measurable levels.

We can write the cross-correlation power spectrum between the surface density fluctuations of galaxies and temperature fluctuations as (for more details, see e.g.~\cite{crittenden96, nolta04, cabre07, raccanelli08})
\begin{equation}\label{eq:ClgT}
C_{\ell}^{gT} = \langle a_{{\ell}m}^g a_{{\ell}m}^{21*} \rangle = 4 \pi
\int \frac{dk}{k} \Delta^2(k) W_{\ell}^g(k)
W_{\ell}^{T}(k) \, ,
\end{equation}
where $W_{\ell}^g$ and $W_{\ell}^{T}$ are the galaxy and temperature window functions respectively, 
and $\Delta^2(k)$ is the logarithmic matter power spectrum today.

The galaxy window function can be written as
\begin{equation}\label{eq:flg}
W_{\ell}^g(k) = \int dz\, \frac{dN}{dz} b(z) D(z) j_{\ell}[k\chi(z)] \, ,
\end{equation}
where  $(dN/dz)dz$ is the mean number of sources per steradian with redshift $z$ within $dz$, brighter than the flux or magnitude limit; $b(z)$ is the bias factor relating the source to the mass overdensity, $D(z)$ is the linear growth factor of mass fluctuations; $j_{\ell}$ is the spherical Bessel function of order $\ell$, and $\chi(z)$ is the comoving distance. We use a redshift distribution and bias for galaxies as predicted for the SKA survey, but our results are not heavily dependent on them.
Selecting a survey with the appropriate median redshift and the bias of the objects observed, one can in principle optimize the detection of the cross-correlation; a detailed study of this is beyond the scope of this paper.

The window function for the ISW effect is
\begin{equation}\label{eq:flT}
W_{\ell}^{T}(k) = \int d \chi \, \, T_{X}(z) \, W(\chi(z),\chi_\nu) \, j_{\ell}[k\chi(z)] \, ,
\end{equation}
where $\chi_\nu$ corresponds to the central frequency targeted by the experiment.

The errors for the cross-correlation can be computed as (see e.g.~\cite{cabre07})
\begin{equation}
\label{eq:err-clgt}
\sigma_{C_{\ell}^{g T}} = \hksqrt{\frac{\left(C_{\ell}^{gT}\right)^2 + \left[ \left( C_{\ell}^{gg} + \bar{n}_g^{-1} \right) \left(C_{\ell}^{TT}+\varepsilon \right)\right]}{(2\ell+1)f_{\rm sky}}} \, ,
\end{equation}
where $f_{\rm sky}$ is the sky coverage of the survey, ${\bar{n}_g}$ is the average number density of sources, and $\varepsilon$ is the instrument noise. In the case of CMB photons this corresponds to the beam noise and it's usually considered to be negligible for modern experiments; for 21-cm experiments we have~\cite{kovetz12}
\begin{equation}
\varepsilon = \frac{2 \pi \beta(z)}{f_{\rm cov}^2 t_0 \Delta \nu } \, ,
\end{equation}
where $\Delta \nu$ is the frequency bandwidth, $\beta(z) = (2\pi)^2 T_{\rm sys}^2(z) (\ell_{\rm cov}^2)^{-1}$, with $T_{\rm sys} \approx 180(\nu/180 {\rm MHz})^{-2.6}$ K, and $\ell_{\rm cov} = 2\pi \mathcal{D}/\lambda$, where $\mathcal{D}$ is the length of the interferometer baseline. Finally, we take $C_\ell^{TT}$ to be the regular CMB power spectrum when considering the signal-to-noise for detection of the ISW effect on the unscattered CMB photons and replace it with the 21cm power spectrum when considering the effect on the scattered ones (assuming perfect CMB subtraction).

In order to detect the ISW effect we need to cross-correlate temperature maps with galaxy catalogs.
In the case of the CMB-galaxy correlation, $\varepsilon\ll C_\ell^{TT}$ and hence the signal-to-noise ratio (SNR) of the cross-correlation depends, as in the ``standard'' ISW case, mostly on the sky coverage of the survey and the number of sources in the galaxy catalog.

The error budget in the case of the scattered photons is dominated by the $\varepsilon$ term of Eq.~(\ref{eq:err-clgt}), so the detectability of this cross-correlation largely depends on being able to reduce it. We study the SNR of the 21-cm-galaxy correlation as a function of the interferometer specifications.

\begin{figure}
\includegraphics[width=0.99\columnwidth]{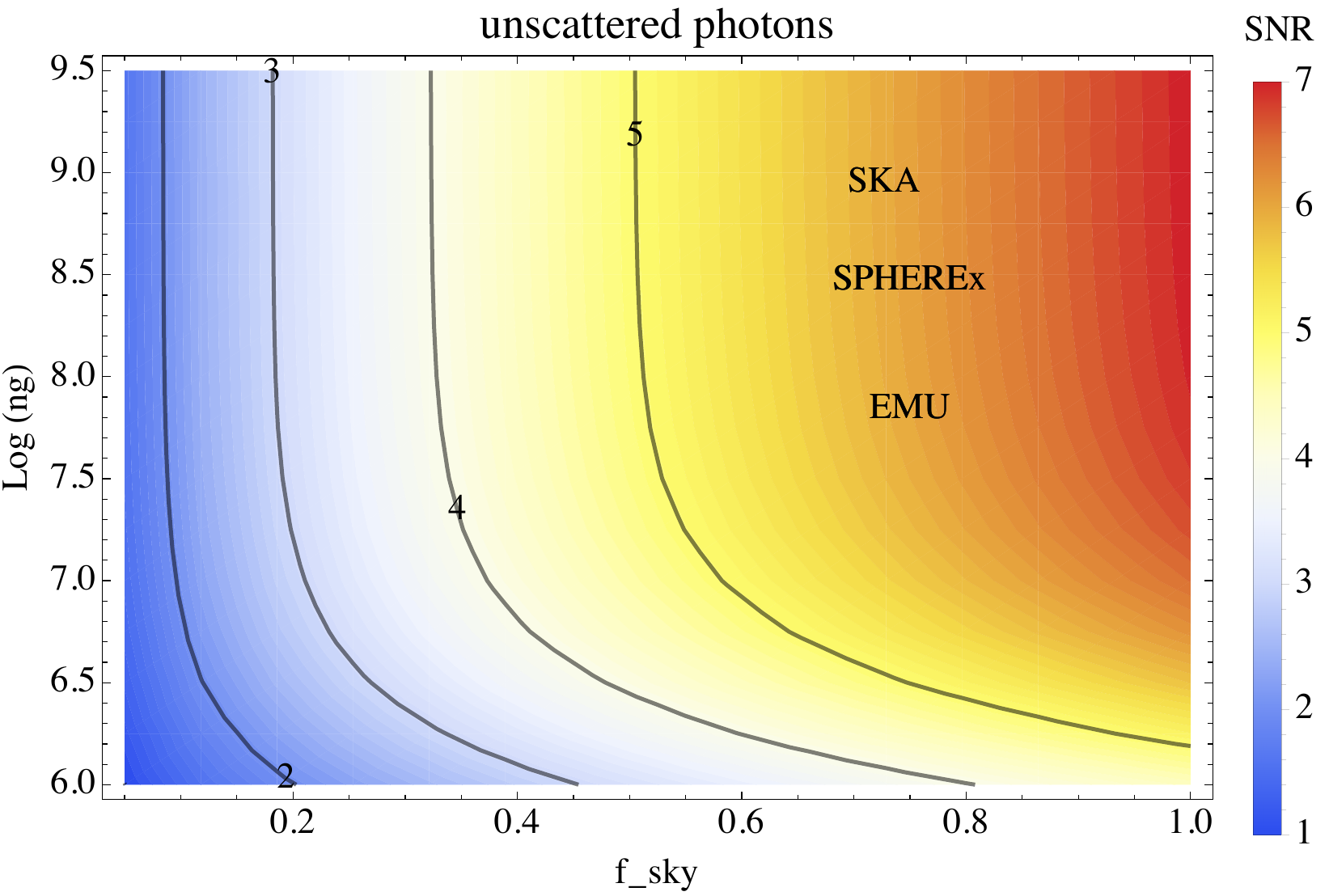}
\includegraphics[width=0.99\columnwidth]{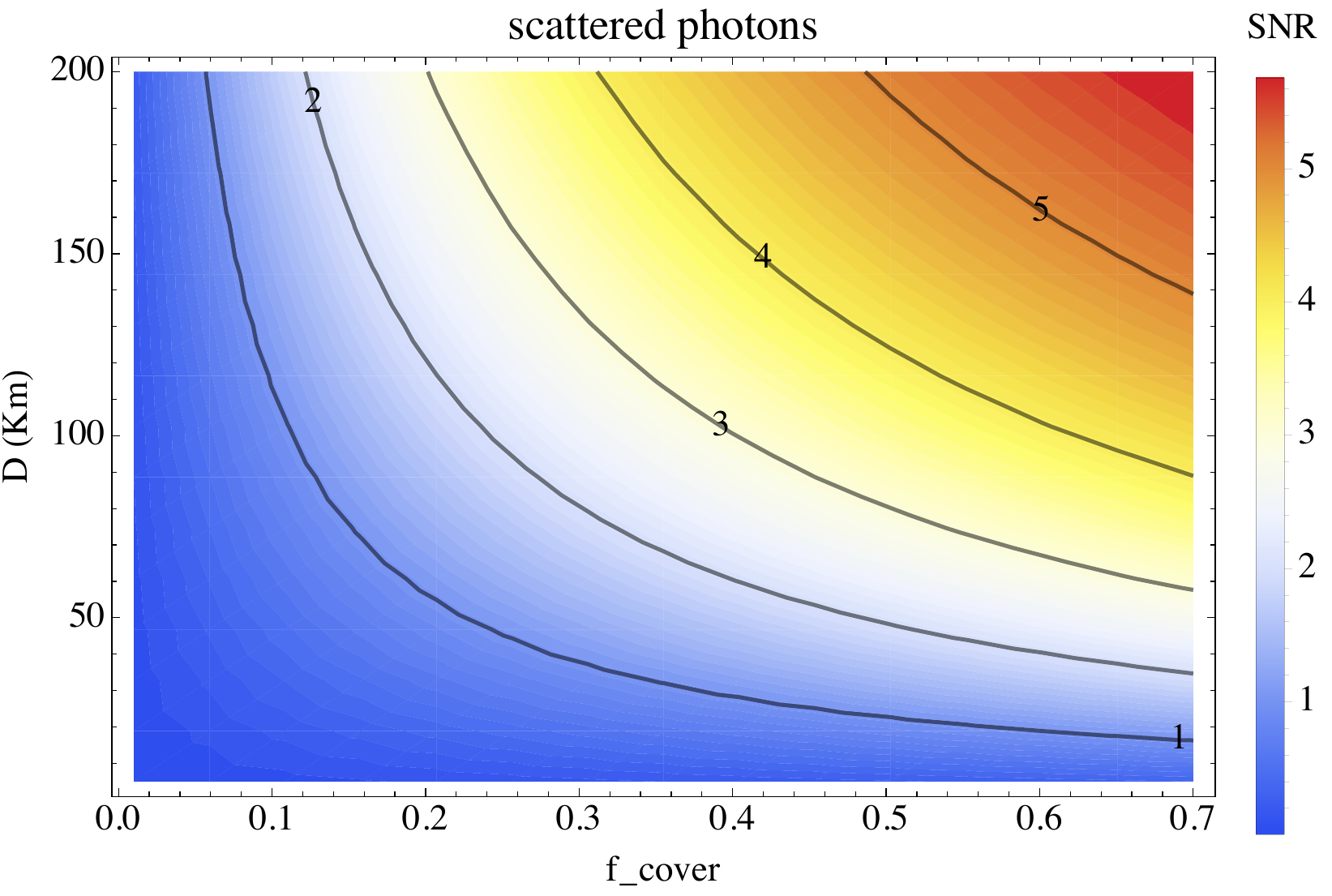}
\caption{{\it Upper Panel}: Signal-to-noise ratio for the correlation of LSS with unscattered photons, as a function of the area of the sky surveyed and the galaxy number density. {\it Lower Panel}: Signal-to-noise ratio for the LSS-21cm correlation, as a function of the fractional area coverage and baseline length.}
\label{fig:snr}
\end{figure}

In Figure~\ref{fig:snr} we show the SNR for the unscattered (upper panel) and scattered (lower panel) cases, where we used the predicted catalog of sources to be detected with the SKA.

In the first case we can see that forthcoming and planned surveys such as ASKAP and the SKA could achieve a SNR of ~5-6 (for specifications of these surveys and predictions about detecting the ISW by cross-correlating their sources with CMB experiments, see~\cite{Raccanelli:2011radio, ASKAP, Raccanelli:2014ISW, SKA:continuum}). Using multiple redshift bins, this signal-to-noise could be further increased by a factor of a few.
In the upper panel of Figure~\ref{fig:snr} we show the SNR for the correlation of LSS with unscattered photons, as a function of the area of the sky surveyed and the galaxy number density.

In the latter case, a detection is, as expected, more difficult to obtain. The $\varepsilon$ term depends on a series of parameters of the 21-cm detecting instrument, such as the observing time, the frequency bandwidth, the fractional area coverage and the length of the baseline. In the lower panel of Figure~\ref{fig:snr} we plot the SNR for the cross-correlation of LSS with 21cm, scattered, photons, as a function of the fractional area coverage and baseline length. To calculate the 21-cm power spectrum, $C_{\ell}^{21\,21}$, we used CAMB Sources \cite{CAMB_sources} with the appropriate frequencies and bandwidths.
The results weakly depend on the galaxy survey used. Different surveys give slightly different results, but we do not find a dramatic change in the overall signal-to-noise ratio; targeting specific redshift ranges and objects could help. Other ways to improve the results could come from the use of a tomographic analysis in the galaxy catalog as well, and from the combination of different surveys (see e.g.~\cite{giannantonio08a, bertacca11}). The combination of different surveys, in particular, has already been shown to improve the detection of the signal in the case of the LSS-CMB correlation.
These optimizations could improve the ability of detecting the ISW from the LSS-21cm correlation, but a detailed analysis of these prospects is beyond the scope of this work and will be studied elsewhere.

To conclude, we have investigated the possibility to detect the ISW effect using high-redshift 21-cm maps measurements. Our results show that currently planned instruments will be able to detect the ISW effect on the CMB photons that pass freely through the neutral hydrogen with a fairly decent sensitivity, allowing a cross-check to those measurements which use the CMB at peak frequencies. As for the effect on the scattered photons, currently planned experiments will not be able to provide a clear detection of the signal, even via the cross-correlation with tracers of the underlying matter distribution such as galaxies.
Future experiments, such as an improved SKA or a lunar radio array on the far side of the moon, however, could possibly grant a signal-to-noise that could allow a clear detection of this cross-correlation. Both measurements could be used as further tests of the standard cosmological model.

\vspace{0.3cm}
{{\bf Acknowledgments}}
\vspace{0.1cm}

This work was supported by NSF Grant No.\ 0244990, NASA
NNX15AB18G, the John Templeton Foundation, and the Simons
Foundation. We would like to thank Yacine Ali-Ha\"{i}moud, Joseph Silk, Jens Chluba, Francisco Villaescusa-Navarro and Stefano Borgani for useful discussions.

\end{document}